\documentclass[twocolumn]{aa}
\usepackage{graphicx}
\usepackage{txfonts}
\begin{document}
   \title{3D spectroscopy with VLT/GIRAFFE - III: Mapping electron
   densities in distant galaxies\thanks{Based on observations
   collected at the European Southern Observatory, Paranal, Chile, ESO
   Nos 71.A-0322(A) and 72.A-0169(A)}}

   \author{M. Puech\inst{1}, H. Flores\inst{1}, F. Hammer\inst{1} \& M.D. Lehnert\inst{2}}

   \titlerunning{Mapping electron densities in distant galaxies}

   \authorrunning{M. Puech et al.}

   \offprints{mathieu.puech@obspm.fr}

   \institute{Laboratoire Galaxies Etoiles Physique et
        Instrumentation, Observatoire de Paris, 5 place Jules Janssen,
        92195 Meudon France 
	\and
	Max-Planck-Institut für extraterrestrische Physik, Giessenbachstrasse, 85748 Garching bei München, Germany
        }

   \date{Received xxx, 2005; accepted xxx, 2005}

   \abstract{We used the moderately high spectral resolution of
     FLAMES/GIRAFFE (R=10000) to derive electron densities from [OII]
     line ratios in 6 galaxies at $z\sim 0.55$. These measurements
     have been done through the GIRAFFE multiple integral field units
     and are the very first \emph{mapping} of electron densities in
     such distant objects. This allows us to confirm an outflow and
     identify the role of giant HII regions in galactic disks. Such
     measurements could be used in the future to investigate the
     nature of the physical processes responsible for the high star
     formations rates observed in galaxies between z$\sim$0.4 and
     z$\sim$1.


}   

   \maketitle

\section{Introduction}

It is now well established that the cosmic star formation density
declines from z $\sim$ 1 to 0 (Lilly et al. 1996; Madau et al. 1998;
Flores et al. 1999; Le Floc'h et al. 2005). However, the physical
processes responsible for this decline are still a matter of debate.
At the heart of this debate is the respective importance of secular
evolution with slow and continuous external matter accretion (e.g.
Semelin \& Combes 2005) versus more violent evolution through
hierarchical merging (e.g. Hammer et al. 2005) as a function of
lookback time. Relating star formation processes with the ISM state
(i.e. its chemistry \emph{and} kinematics) of distant galaxies could
provide a new insight into this debate. We present in this paper the
first mapping of electron density in a small sample of distant
galaxies. Our goal is to demonstrate the feasibility of such
measurements using current integral field spectrographs on a 8 meter
telescope. In the future, we will investigate the possible relations
between such quantities (e.g. electron density, metal abundance ...)
and the star formation rate. Their possible correlations could help in
constraining the physical processes at work in galaxies, those which
are responsible for the strong evolution of the cosmic star formation
density.

Electron density can be determined from the intensity ratio of two
lines of the same ion arising from levels with nearly the same
excitation energy (Osterbrock \& Ferland 2006). The lines usually used
are the [OII]$\lambda \lambda$3729,3726\AA~and the [SII]$\lambda
\lambda$6716,6731\AA~doublets. For distant galaxies, the [OII] doublet
becomes particularly well suited to probe electron densities, although
local [OII] variations in galaxies are more sensitive to extinction
and metallicity (e.g. Kewley et al. 2004; Mouhcine et al. 2005). First
determination of electron density $N_e$ from [OII]$\lambda
\lambda$3727,3729$\AA$ line ratio $r=3729/3727$ in nebulae has been
suggested by Aller et al. (1949). The relation between $r$, $N_e$ and
electron temperature $T_e$ has been successively improved by Seaton et
al. (1954) and Eissner et al. (1969). First confrontations with
observations were done in several nearby nebulae with good agreement
(Seaton \& Osterbrock 1957). Because $r$ only depends weakly on $T_e$
(see e.g. Canto et al. 1980), this relation provides a good mean to
measure $N_e$ and then investigate the structure of HII regions such
as Orion (Osterbrock et al. 1959; Danks et al. 1971; Canto et al.
1980) or M8 (Meaburn et al. 1969), and of planetary nebulae (see
Osterbrock \& Ferland 2006 and references therein). More recent
studies of several HII regions have been carried out by Copetti et al.
(2000) in our Galaxy and by Castaneda et al. (1992) in the local
group.

We recently used the FLAMES/GIRAFFE spectrograph at ESO/VLT in its
multi-IFU mode to target the [OII] doublet of 35 galaxies at $z \sim
0.55$ in order to obtain their kinematics (see Paper I, Flores et al.
2006 and Paper II, Puech et al. 2006a). Thanks to the excellent
spectral resolution of GIRAFFE (R $\sim$ 10000), we have used here a
part of this sample to derive the electron density maps of a few
distant galaxies. This paper is organized as follows: section 2
presents the sample and the methodology, results are in section 3 and
section 4 gives a conclusion.

\section{Sample, observations \& methodology}
As part of the GTO of the Paris Observatory, we obtained
FLAMES/GIRAFFE-IFU (3 by 2 arcsec$^2$, 0.52 arcsec/pixel) observations
for 35 galaxies at $0.4 \leq z \leq 0.71$ in the CFRS (03hr and 22hr)
and HDFS field. The complete sample is described in Paper I (Flores et
al. 2006) and the Luminous Compact Galaxies (LCGs) sub-sample in Paper
II (Puech et al. 2006a). Briefly, we used LR04 and LR05 setups
targeting the [OII] doublet (R \(\sim10000\)) with integration times
ranging from 8 to 13 hours. Seeing was typically \(\sim 0.6\) arcsec
during all the observations. Datacubes were reduced using the GIRBLDRS
v1.12 package (Blecha et al. 2000), including narrow flat-fielding,
and the sky was carefully subtracted using our own IDL procedures.

Among these 35 galaxies, we selected those for which at least 1.5
spatial resolution element had [OII] doublets reaching a mean SNR per
spectral resolution element of 8 (see Paper I). This allows the
establishment of meaningful electron density maps, i.e. with at least
6 pixels among the 20 composing the GIRAFFE IFU. For galaxies with
complex kinematics, the large pixel size of GIRAFFE integrates both
random motions and larger scale motions. This tends to blend the [OII]
doublet in these kinds of galaxies in spite of the high spectral
resolution of GIRAFFE ($\sim$ 10000, see Paper I, Flores et al. 2006).
For the present study, we discarded all galaxies where this effect
could lead to a too high uncertainty on the [OII] line ratio
measurement. We finally selected 6 galaxies (see table \ref{tab1})
among those having the highest quality factor of the whole sample
(Paper I, Flores et al. 2006).

\begin{table}[h!]
\centering
\begin{tabular}{cccc}\hline\hline
ID         & z      & I$_{AB}$ &  M$_B$\\\hline
CFRS03.0488    & 0.6069 & 21.58      &  -20.37 \\
CFRS03.0508    & 0.4642 & 21.92      &  -19.61 \\
CFRS03.0645    & 0.5275 & 21.36      &  -20.30 \\
CFRS03.9003    & 0.6189 & 20.77      &  -21.24 \\
CFRS22.0504    & 0.5379 & 21.02      &  -20.52 \\
CFRS22.0919    & 0.4738 & 21.77      &  -19.99 \\\hline\hline
\end{tabular}
\caption{Main properties of the sample of galaxies: galaxy names,
redshifts, isophotal I magnitude and absolute B magnitude (from Hammer
et al. 2005).}
\label{tab1}
\end{table}

[OII] doublets were fitted after a slight Savitzky-Golay filtering
which has the advantage of conserving the first moments of the
spectral lines (Press et al. 1989). During the fit we used the
following constraints: $\lambda _2$-$\lambda _1$=2.783$\AA$ and
$\sigma_1$=$\sigma _2$ (see Paper I, Flores et al. 2006 and Paper II,
Puech et al. 2006a for the detailed procedure). We checked by visual
inspection each fit and discarded a total of 4 pixels ($\le$ 1 \% of
pixels) where the results were particularly uncertain, mostly due to
noisy peaks or extinction effects. We then derived [OII] distribution
maps by simply integrating the fitted doublets in each pixel. [OII]
line ratio was related to electron density using n-levels atom
calculations of the stsdas/Temden IRAF task. The relation linking $r$,
$N_e$ and $T_e$ depends only weakly on $T_e$ in the range of
temperatures of the regions studied (see Eissner et al. 1969, Canto et
al. 1980): we took $T_e = 10000 K$ in all the sample, which is a good
approximation for most of HII regions (Osterbrock \& Ferland 2006).
The maximal value of the line ratio $r$ in this calibration is then
1.492. Using some different collision strengths from Mendoza (1983),
one can derive a maximal ratio of 1.497 (J. Walsh, private
communication). Given our uncertainty on the measurement of $r$ (0.05,
see below), one can derive an upper limit of 1.56 for acceptable
measures of $r$: we checked that all the measured line ratio were
lower than this limit, and forced all line ratio greater than 1.492
(and lower than 1.56) to 1.492 (the last point in the calibration).
This corresponds to a density of $\sim$ 1 $cm^{-3}$ and affects $\sim$
8 \% of the pixels.

The main sources of errors in the determination of electron density
are twofold. The first one is the error on the determination of $r$
during the fit: compared to line positions and widths, [OII] line
ratios are the less well determined parameters of the fit because it
is more sensitive to noise. We estimated this error to be typically
$\sim$ 0.05 (3$\sigma$) on $r$. For example, for a ratio $r=1.3$, one
derives $N=101^{+33}_{-29} cm^{-3}$. The second one is due to the
saturation of the $r$ vs $N_e$ relation at low densities (typically at
$N_e \lesssim10 cm^{-3}$) and is very difficult to estimate. As
already mentioned above, we tried to minimize these effects by
limiting the measurement of $r$ to the highest SNR pixels reaching a
mean SNR per spectral resolution element greater than 8 which
corresponds to a higher threshold than the one used for the other maps
(see Paper I, Flores et al. 2006). Finally, the main uncertainty
results from extinction effects: As the [OII] emission line can be
severely affected by extinction (see e.g. Hammer et al. 2005), some
local density peaks could be hidden by dust and then undetected by the
[OII] line ratio diagnostic.

Finally, given the GIRAFFE IFU spatial resolution (20 pixels), we
interpolated all maps (velocity fields, $\sigma$-maps, electron
density maps and [OII] fluxes maps) by a simple 5x5 linear
interpolation to make visualization easier (see figure \ref{Fig1}).

\section{Results}

\begin{description}

\item{\bf {CFRS03.0488}} This galaxy has an asymmetric morphology
which is usually believed to be a signature of interactions and/or gas
accretion. Its kinematics is classified as complex by Paper I (Flores
et al. 2006). The [OII] map shows a peak off centered relative to the
brighter component of the galaxy, i.e. in the diffuse component (above
the main optical component), apparently correlated with a higher
electron density region. Densities are in the range [30-153] $cm^{-3}$
which are typical of classical HII regions (Copetti et al. 2000). The
higher densities observed in the diffuse component could be explained
by collisions between molecular clouds of the interstellar medium and
gas inflow/outflow events, which are suggested by the morphology.\\

\item{\bf {CFRS03.0508}} This Luminous Compact Galaxy (LCG) was
classified by Zheng et al (2005) as a relic of an interaction or
fusion, with a relatively blue color over the whole galaxy which is
also seen in the [OII] map. In Paper II (Puech et al. 2006a), we found
that its kinematics show a rotational motion pattern with an axis
almost perpendicular to the main optical axis of the galaxy, which
probably reflects an outflow motion rather than a rotational motion.
Electron densities are characteristic of classical HII region ($\le$
100 cm$^{-3}$), higher concentrations being well aligned along the
supposed outflow (maximal densities are pointing to maximal velocities
ends). We believe this supports the outflow hypothesis, because in
such a case, electrons are produced from collisions between the
expelled gas and molecular clouds of the interstellar medium.\\

\item{{\bf CFRS03.0645}} Zheng et al. (2005) found in this LCG a
relatively blue color all over the galaxy and classified it as a
probable merger. The [OII] map shows a peak in the central region. As
for CFRS03.0508, we found in Paper II (Puech et al. 2006a) a
kinematics showing a rotational motion pattern with an axis almost
perpendicular to the main optical axis of the galaxy, which could
reflect an outflow motion rather than a rotational motion. The
$\sigma$-map has a peak at the edge of the galaxy, where the two main
components join. Electron densities are characteristic of classical
HII regions ($\sim$ 100 cm$^{-3}$) and a peak at $\sim$ 200 cm$^{-3}$
is found near the main optical component (at the upper right of the
image). Unfortunately, the SNR was too low to map the diffuse
component at the bottom of the HST image, and thus we cannot confirm
the suspected outflow. \\

\item{{\bf CFRS03.9003}} This Luminous IR Galaxy (LIRG, SFR $\sim$ 75
$M_{\odot}/yr$, see Flores et al. 2004) presents a quite complex and
irregular brightness distribution. Its kinematics is classified as a
rotating disk by Paper I (Flores et al. 2006). Zheng et al. (2004)
found blue star-forming regions surrounding a red central region that
might be a bulge. The [OII] counts distribution is relatively flat
(between $\sim$ 2 and 9) but present a small relative maximum (on the
right of the image) which is located in the disk. The central region
has a very low density (i.e. a few $cm^{-3}$) comparable with those of
free electrons in the interstellar gas outside of HII regions
(Reynolds 1989): this corresponds to a gas poor region and it is
consistent with the presence of a bulge in this galaxy. The off
centered knot in the I band HST image corresponds to a blue region
with a peak in electron density. Paper I (Flores et al. 2006) showed
that this knot is not due to a minor merger since no perturbation was
seen in its kinematics (see figure \ref{Fig1}). We found a
significantly higher density in this region (almost 400 cm$^{-3}$),
comparable with those found in the outer areas of some nearby nebulae,
such as M8 or Orion (Meaburn 1969; Osterbrock et al. 1959). This
relatively high density could, at least partly, contribute to the very
high SFR observed in this LIRG. The knot observed in the HST image is
thus probably an extremely large HII region ($\sim$ 2 kpc, a complex
of several HII region ?) undergoing a strong star formation episode.\\

\item{{\bf CFRS22.0504}} Its kinematics is classified by Paper I
  (Flores et al. 2006) as rotating disk. Electron densities are higher
  near the edges. The [OII] map shows maxima near the center and also
  in a region near the edge where we can also see a peak (67
  cm$^{-3}$) in the electron density map. This probably indicates a
  star forming region in the disk. Unfortunately, we only had at our
  disposal a (deconvolved) CFHT image (0.207 arcsec/pix) to be used
  for comparison.\\

\item{{\bf CFRS22.0919}} The morphology of this LCG shows two tails on
each side, characteristic of interacting systems. Its kinematics is
classified complex by Paper II (Puech et al. 2006a) which further
supports a merger hypothesis. Electron densities peak in the tails
($\ge$ 200 cm$^{-3}$) which corresponds to what is expected in
interacting systems.\\
\end{description}

\begin{figure*}[p]
\centering
\caption{SEE ATTACHED JPEG FIGURE Mapping of 6 distant galaxies. All maps have a scale of
$\sim$0.1 arcsec/pix and show the same FoV (3x2 arcsec$^2$). From left
to right: I band HST imaging except CFRS220504 (deconvolved CFHT
image, see text); velocity field (5x5 interpolation), $\sigma$-map
(5x5 interpolation), electron density map (5x5 interpolation), and
[OII] counts distribution map (5x5 interpolation). Electron density
maps have been restricted in SNR (see text).}
\label{Fig1}
\end{figure*}

\section{Discussion \& conclusion}
We measured [OII] doublet lines ratio of 6 galaxies with high SNR to
derive the first mapping of electron density in z$\sim$ 0.6 galaxies.
The sample of the 6 objects presented here includes a large variety of
objects with obvious merger (CFRS22.0919), suspected outflows
(CFRS03.0508 and CFRS03.0645), and spiral galaxies (CFRS22.0504)
including a LIRG (CFRS03.9003). Such a mapping can be very powerful
for understanding the physical processes at work in these galaxies.
The most spectacular illustrations are CFRS03.0508 where an outflow
has been confirmed and CFRS03.9003 where a giant HII region has been
identified. We also derive maps of [OII] total counts. For the 3 LCGs
of the sample, we find that the [OII] maps show a peak at their
centers, corresponding to the relatively blue cores found by Zheng et
al. (2005). For the 2 rotating disks (CFRS03.9003 and CFRS22.0504),
[OII] counts are distributed over the disks. This confirms that star
formation migrates from center to the outskirts of the disk when
comparing LCGs to spirals (Zheng et al. 2005).

The main limitation of our results arises from the fact that GIRAFFE
has a relatively low spatial resolution (0.52 arcsec/pix): this makes
the derived electron densities underestimated because they are
averaged on spatial regions bigger than the characteristic length of
HII regions ($\sim$ 100 pc). Another consequence of the low spatial
resolution of GIRAFFE is that line widths are the convolution of
random motions with larger scale motions. In the case of merging
systems where velocities can be particularly high (see Paper I, Flores
et al. 2006), this makes the [OII] doublet blended in spite of the
high spectral resolution of GIRAFFE (R$\sim$10000) and of the SNR
level, and then only velocities and $\sigma$ can safely be recovered.
Mapping such galaxies with complex kinematics are currently beyond the
capabilities of an Integral Field Spectroscopy on 8 meter telescopes
because it would require a much higher spectral and spatial
resolution. Finally, another important limitation can be due to
extinction by dust which could hide some density peaks: only the
unobscured regions can contributed to the [OII] flux detected, and the
density maps are thus biased toward these regions.

With the arrival of Integral Field Spectrograph operating in the NIR
(such as SINFONI), it is now possible to extend this kind of mapping
to other physico-chemical parameters such as extinction, instantaneous
SFR and metal abundance. The investigations of possible correlations
between these quantities would probably shed a new light in the
processes leading to the intense star formation rates observed at
z$\geq 0.4$ and the decrease of the cosmic star formation density
since z$\sim$1.

\begin{acknowledgements}
We wish to thank ESO Paranal staff for their reception and their very
useful advises during observations. We also thank A. Rawat for
improving the English of this paper and the referee, Pr. J.M. Vilchez,
for comments and suggestions. We thank all the team of GIRAFFE at
Paris Observatory, at Geneve Observatory and at ESO for the remarkable
accomplishment of this unique instrument, without which, none of these
results would be obtained.
\end{acknowledgements}


\end{document}